\documentclass[useAMS,usenatbib]{mn2e}
\usepackage{epsfig}
\title[Do dwarf galaxies form in tidal tails?]{Do dwarf galaxies form in tidal tails?}

\author[M. Wetzstein, T. Naab and A. Burkert]{M. Wetzstein$^{1}$\thanks{E-mail: mwetz@usm.lmu.de}, T. Naab$^{1}$and A. Burkert$^{1}$\\
$^{1}$Universit\"ats-Sternwarte, Scheinerstr. 1, D-81679 M\"unchen, Germany}

\begin{document}

\newcommand{\apj}{ApJ} 
\newcommand{\apjs}{ApJS} 
\newcommand{\apjl}{AJL} 
\newcommand{\aj}{AJ} 
\newcommand{\mnras}{MNRAS} 
\newcommand{\aap}{A\&A}
\newcommand{\aaps}{A\&AS}
\newcommand{\araa}{ARA\&A}
\newcommand{\nat}{Nat}
\newcommand{\pasj}{PASJ}
\newcommand{\pasp}{PASP}
\newcommand{\apss}{Astrophys. Space Sci.}
\newcommand{\prevl}{PhysRevL}
\newcommand{\jcomp}{JCP}
\newcommand{\cpr}{Comp. Phys. Rep.}
\newcommand{\cpc}{Comp. Phys. Comm.}
\newcommand{\siamcomp}{SIAM J. Sci. Comp.}
\newcommand{\siamstat}{SIAM J. Sci. Stat. Comp.}
\newcommand{\memras}{MmRAS}
\newcommand{\baas}{BAAS} 

\newcommand{\m}{\mathbf}
\newcommand{\f}{\frac}
\newcommand{\ov}{\overline}
\newcommand{\p}{\partial}
\newcommand{\mr}{\mathrm}
\newcommand{\tb}{\textbf}
\newcommand{\tx}{\texttt}
\newcommand{\ti}{\textit}
\newcommand{\beq}{\begin{equation}}
\newcommand{\eeq}{\end{equation}}

\pagerange{\pageref{firstpage}--\pageref{lastpage}} \pubyear{2005}

\maketitle

\label{firstpage}

\begin{abstract}

The formation of tidal dwarf galaxies (TDG) inside tidal arms of 
interacting disk galaxies has been studied with \ti{N}-body and 
\ti{N}-body/SPH simulations at different resolutions. In pure \ti{N}-body
simulations no bound objects are formed at high resolution. 
At low resolution bound objects can form in tidal tails in agreement with
previous work. We conclude that tidal dwarf galaxies are not likely to form by
pure collisionless collapse in tidal tails. However, the presence of a
sufficiently massive and extended gas component in the progenitor disk
supports the formation of bound stellar objects in the tidal arms. Our results
clearly favor a dissipation supported scenario in which the formation  of TDGs
is induced by the local collapse of gas which then triggers the collapse of
the stellar component. 
\end{abstract}


\section{Introduction}

\subsection{Observations of Tidal Tails in Interacting Galaxies}
\label{tdg_obs}

Interacting galaxies are very complex and highly dynamic systems. They
can be comprised of spiral galaxies as well as elliptical galaxies which
possibly merge and change morphological type. If the mass ratio of spiral
mergers is in the range $1:1$ to $4:1$, from theoretical as well as
observational point of view the systems are likely to evolve into elliptical
galaxies \citep[][ but see \citealp{robertson2006} for a counter
example]{genzel2001}. For those mass ratios the remnant properties 
resemble elliptical galaxies \citep{naab2003, naab&trujilo2006}. Typically,
mergers take place in 
field and group environments where the velocity dispersion is of the
order of the escape speed of the galaxies. However, merging has also
been observed in probably unrelaxed clusters
\citep{vandokkum1999,tran2005}. 

Mergers between spheroidal galaxies are not spectacular. As these systems are
mostly gas poor and dynamically hot the merger is neither accompanied by a
starburst nor by distinct tidal features \citep{vandokkum1999, bell2006,
  naab2006a}. Hence the formation of
new dwarf galaxies inside tidal features is very unlikely.

If spiral galaxies with gas and dynamically cold discs are involved in
a merger the systems show striking features like starbursts, AGN
activity \citep{genzel1998, sanders1996,dasyra2006} and long arms of
stellar and gaseous material which extend from the parent galaxy to sometimes
hundreds of kiloparsecs away. Some famous examples are NGC
4676 ('The Mice'), NGC 4038/39 ('The Antennae'), IRAS 19254-7245 ('The
Superantennae') \citep[see e.g.][ and references therein]{hibbard1996,
  wilson2000, hibbard2001, fabbiano2003, hibbard2005, mirabel1991, vanzi2002,
  charm2002}.
These arms are produced during the close encounters of the  
galaxies when strong tidal forces expel the material from the discs of
their parent galaxies \citep{toomre1972}. Already \citet{zwicky1956} suggested
that inside these tidal tails new stellar systems could be
forming. This idea was revived by \citet{schweizer78} who found a low surface 
brightness object slightly beyond the tip of the southern tail in NGC
4038/39. Since then, a lot effort has been spent
on observations of candidate objects for such tidal dwarf galaxies (TDGs).

The distribution of gas and stars in tidal tails can be very different
  \citep{hibbard1996, hibbard1999, heith2000, smith1997,
  hibbard2000}. Some tails show gaps which could 
be a consequence of local collapse or feedback associated with star formation
\citep{hibbard1996} and in some tails, there is an offset between the
stellar and the gaseous material \citep{hibbard1999}. Tidal tails can easily
  contain up to several $10^9$ 
M$_\odot$ of gas and a similar mass in stars \citep[see
  e.g.][]{hibbard1994,hibbard1999,hibbard2001}. The molecular complexes
  forming in the tails can reach masses of a few $10^7$ M$_\odot$ \citep[see e.g.][]{heith2000,walter1999}.

The formation processes of TDGs and star clusters in tidal tails seem to be
linked, as tidal tails with TDG candidates seem to have few star clusters,
while those with many star clusters have no TDG candidates \citep{knierm2003}.
Both star cluster formation as well as TDG formation are
directly coupled to the local properties of the gas
reservoir. The formation of TDGs might rely on exceeding a threshold in HI
column density \citep{duc2000}, much like the similar
threshold for star formation in dwarf irregulars and low surface brightness
galaxies \citep{davies1976, gall1984, skillm1987, taylor1994,
  hulst1993, zee1997}. As the threshold should be metallicity dependent
\citep{franco1986}, TDGs should have a lower threshold than dwarf galaxies
formed in isolation because TDGs are made from recycled material of the progenitor discs \citep{duc2000}. 

To the present day, there are around 60 well known TDG candidate objects
\citep{schweizer78, mirabel1992, hibbard1994, yoshida1994,
  duc1997, duc1998, heith2000, duc2000, oliveira2001, weilb2002,
  temporin2003b, weilb2003, mundell2004, bournaud2004, amram2004}. However,
the quality and level of details of the observations still varies considerably
for these objects, which also introduces some scatter in the reliability of
their detections as TDG candidates. 

In general a TDG can be detected either while it is still embedded in the
tidal tail or as a dwarf galaxy accompanying an already merged system. In the
first case, one will try to detect overdensities in gas and stars inside the
tidal tail, with the ultimate proof for a newly formed TDG being that
it kinematically decoupled from the surrounding tail
\citep{duc2000}. However, corresponding signatures in velocity
gradient or dispersion can easily be masked by projection
effects \citep{bournaud2004} and so far only only one such object has
been confirmed \citep{bournaud2004}, while for other observed objects
the velocity gradients are not large enough to assure self-gravity \citep[see
e.g.][]{oliveira2001, hibbard2001,saviane2004, amram2004, bournaud2004}. 

Alternatively, one can detect TDGs after their
decoupling from the tidal tail, i.e. to search the population of
dwarf galaxies in merger systems for TDGs \citep[see e.g.][]{deeg1998,
  delgado2003}. \citet{duc1998} suggested to use a metallicity based selection
criterion of $\approx$Z$_\odot/3$ for TDGs. Given the higher metallicity of
TDGs compared to usual dwarf galaxies, TDGs would thus also deviate from the
luminosity-metallicity relation of dwarf and giant galaxies \citep[see
e.g.][]{skillm1989, richer1995, skillm1997}.

Recently, a sample of 13 TDG candidate objects has become available \citep{weilb2000, weilb2002, weilb2003}. The majority of these TDG candidates has
luminosities of $M_B \approx -12$ to $-17$ in $B$-band and a mean
H$\alpha$ luminosity of 2.2 $\times 10^{39}$ erg s$^{-1}$. Their mean
oxygen abundance is $12+\log(O/H)=8.34 \pm 0.2$, consistent with
material in the outer disc of spiral galaxies like the Milky Way
\citep[see e.g.][]{naab_ostrik2006}. All 13 objects seem to have
internal velocity gradients, but higher resolution spectra would be
required for confirmation.

\subsection{Theoretical Modelling of Tidal Tails} 
\label{tdg_theo}

The formation and evolution of tidal features in systems of
interacting galaxies has been studied with theoretical models since the
early 1960s \citep[see e.g.][]{pfleiderer1961, pfleiderer1963, 
  toomre1972, yabushita1971, clutton72, clutton1972b, wright1972,
  eneev1973}. Most of these models relied on restricted three-body
simulations. With the advent of more powerful computers, also more 
advanced algorithms for such particle simulations became available. Tree codes
have been the method of choice for many merger studies, as they offer a wide
dynamic range regarding their spatial and mass resolution \citep[see
e.g.][]{appel85, jernigan85, barnes_hut86, press86, jernigan_porter89,
  barnes90}. Gas has been incorporated into such tree code simulations using
the Smoothed Particle Hydrodynamics method, or short SPH \citep{lucy77,
  ging_mon77,benz90,monaghan92}.

Self-consistent merger simulations have been used since the mid 1980s to study
the formation of the remnant galaxy and 
its properties. General studies of merging systems \citep[see
e.g.][]{barnes1988, barnes1988b, barnes1989, naab1999, naab2003} have been
conducted as well as attempts to model specific observed merger systems
\citep[see e.g.][]{mihos1993, hibbard1995, mihos1997, mihos1998, duc2000,
  barnes2004}. The distribution of gas and stellar material
in the tidal tails is of major importance to determine the encounter geometry
of the observed system. However, those simulations reproduced the tidal tails
as such and were not particularly aimed at modelling the formation of bound
structures in the tidal tails. 

Reproducing the star formation history of merging systems has been another
constraint for simulations. However, the star formation rate can easily be
underestimated in regions where strong compression due to shocks in the
encounter occur, which also makes it difficult to model merger induced
starbursts  which are strong enough to yield emission comparable to observed
ULIRGs \citep{mihos1993}. More refined star formation prescriptions in the
models are required to model such star bursts \citep{barnes2004}.

Merger simulations have also been successful in understanding the offsets
between gas and stellar material observed in some merging systems
\citep{smith1997, hibbard1999, hibbard2000}. Rather than being due to complex
feedback phenomena in the tails, those offsets are due to dissipational
effects  during the interaction, which lead to a kinematical decoupling of the
gaseous component from the stellar one soon after the tidal tails are
ejected \citep{mihos2001}. Hence such offsets are a kinematical consequence
of some specific encounter geometries. 

The tidal debris of a merger can be used to probe the properties of the
underlying dark
matter halos \citep{faber1978,white1982,negrop1983,barnes1988}. Merger models
in which the dark matter halos are set up as pseudo-isothermal spheres
\citep[see e.g. ][]{hernquist90} or NFW halos \citep{nfw96,nfw97} can form
significant tidal tails, provided that the ratio of escape velocity $v_e$ to
circular velocity $v_c$ at about the solar radius in the
progenitor galaxy is $v_e/v_c \le 2.5$ \citep{dubinski1999, springel1999}. 

Most merger simulations concentrated either on the properties of the remnant
galaxies or on the global distribution of mass in tidal
tails. \citet{barnes_hern1992}  were the first to concentrate explicitly on the
formation of structures inside the tidal tails. They modeled one encounter,
corresponding to one pair of tidal tails, in which they found $23$ clumps of
material, possibly being TDGs or their progenitors. The resolution of 
these tails was limited by the low numbers of particles in the simulation,
in total $\approx 45000$ per galaxy with $\approx 8000$ of them for the gas
component. The 
most massive clump of material in the tidal tails was resolved with 350
particles and had a total mass of $4.18 \times 10^8$ M$_\odot$ if the
progenitor galaxies are scaled to the 
Milky Way \citep{barnes_hern1992}. 25\% of the mass inside the object was in
gas. The dark matter content of the object was $<5\%$. The distance from the
remnant at which the most massive object forms varies between pure N-body
simulations of the same encounter \citep{barnes1992} and simulations including
gas \citep{barnes_hern1992, barnes_h1996}. While the stellar material which
ends up in the TDG candidates gently collapses once removed from the parent
galaxy, the gas is driven into a thin 'ridge line' along the tail by convergent
flows. The bound objects were thought to form in the stellar component alone,
then gas 
assembles in these potential wells \citep{barnes_hern1992, barnes_h1996}. This
formation scenario has been one of the two standard paradigms for the
formation of TDGs. \citet{barnes_h1996} emphasize that their pure $N$-body
simulation showed similar objects than the one including gas and thus point out
that ``dissipative effects  are not crucial in forming such structures'', as
the density fluctuations in the stellar component alone were sufficient to
produce the TDG candidates. This should, however, be taken with care, as the
source for the density fluctuations in the tail which then collapse and form
TDG candidate objects are most probably particle noise and swing amplification
during the merger \citep{barnes_hern1992}, which means that at different
resolutions and thus different particle noise levels, the fragmentation of a
tidal tail possibly looks quite different which hasn't been addressed since
\citet{barnes_hern1992}. We will attempt to do so in section \ref{sec_res}.

\citet{elmegr1993} and \citet{kaufman1994} suggested a different process for
the formation of bound clumps in tidal arms. This scenario, in which gas plays
the key role in the formation of TDGs is the other major paradigm established
for the formation of TDGs. The simulations of \citet{elmegr1993} and
\citet{kaufman1994} show the growth of large cloud complexes as massive as
$\approx 10^8$ M$_\odot$ in the tidal arms. The contribution of gas to the
total mass of those objects was $\approx 60\%$. \citet{elmegr1993} explained
the formation of 
these objects with the high velocity dispersion in the gas of interacting
systems. A cloud forming from gravitational instabilities inside an ISM with
increased velocity dispersion is likely to be more massive, as the local Jeans
mass of the gas is increased as well. In addition, the dispersion intrinsic to
the cloud should also be increased, allowing it to form stars with higher
efficiency as the cloud is more stable against self-destruction. The
conclusion of \citet{elmegr1993} and \citet{kaufman1994} that gas is the key
ingredient for the formation of TDGs clearly contradicts the results of
\citet{barnes_hern1992}. It should be noted that the numerical models used to
arrive at these two results differ considerably in terms of the numerical
methods as well as the initial conditions.

\citet{li2004} simulated the formation of
globular clusters in tidal tails. Although they were able to reproduce the
distribution and metallicity of globular clusters in ellipticals, the insight
on TDG formation in those simulations is rather limited because of the use of
sink particles \citep{bate1995} for modelling the collapse of gaseous
structures into globular clusters. As every sink particle resembles one
globular cluster, no larger stellar dynamical systems like TDG are modelled. 
Stellar superclusters formed in mergers possibly merge later and form
ultracompact dwarf galaxies \citep{kroupa1998,fellh2002}.

Recently there has been some discussion about the conditions under which TDGs
can be formed at the very tip of the tidal tail, a location in which they or
their progenitors are frequently found in observed systems (note that not
all TDG candidates are located at the tip of their ambient tidal
tail, though). \citet{bournaud2003} and \citet{duc2004} have argued that there
could exist two distinct formation mechanisms for TDG candidates: Local
gravitational instabilities in the inner part of tidal tails lead to the
formation of TDG candidates with masses of $10^7-10^8 M_\odot$ while massive
($> 10^9 M_\odot$) TDG candidates at the very tip of tidal tails form in a
top down scenario in which the tidally expelled gas accumulates at the tip of
the tail and then undergoes gravitational collapse. This process might
be similar to the process of collapsing sheets
\citep{burkert2004b}. \citet{bournaud2003} and 
\citet{duc2004} argue for a threshold in the radial extension of the
progenitor galaxy's dark matter halo relative to the size of the progenitor's
stellar disc. Only haloes with cutoff radii at least $10$ times larger than the
radial scale length of the disc should hence be able to form such massive TDG
candidates at the tip of tidal tails.

However, some care is advised in the interpretation of those
simulations, as they have a rather coarse spatial resolution of the $N$-body
component of $5$ kpc, which
is why the gravitation of the gas component in \citet{bournaud2003} 
is only simulated in two dimensions, while the rest of the simulation
is three dimensional. The 2D resolution is $150$ pc at the cost of neglecting
the extent of the gas in the tail perpendicular to the orbital plane
completely. This 
problem is overcome in \citet{duc2004}, but still the resolution is only $390$
pc in one simulation and even a factor of two worse for all other
simulations. Both studies don't model the gas component with full
hydrodynamics, but with the simpler sticky particles approach.

In this paper we explore the origin of TDGs with high resolution
simulations. In Section \ref{sec_ic} we
describe the initial conditions of our numerical models and in Section
\ref{sec_params} the simulation code. The gravitational force
softening used in our simulations is discussed in Section
\ref{sec_soft_setup}. The results of our simulations are presented in
Section \ref{sec_res}, where we discuss in detail the influence of
numerical resolution on structure formation in the tidal tails in
Section \ref{sec_totres}. The effects of gravitational
softening on the results are discussed in Section
\ref{sec_soft_result}, while the effects of gas in the tails are
discussed in Section \ref{sec_gasinf}. In Section \ref{sec_tdg} we present
detailed photometric and kinematical analysis of our most massive TDG
candidate. Finally, we summarize and conclude in Section \ref{sec_conclusion}.


\section{NUMERICAL MODELS}
\subsection{Initial Conditions}
\label{sec_ic}

Every galaxy in our simulations consists of an exponential stellar
disc component, a stellar bulge and a dark matter halo. In addition, a
subset of the simulations included also a gaseous disc. The galaxies have been
set up with the method described by \citet{hern1993} and have been run in
isolation to verify that they are in equilibrium for a time scale longer than
the time until the first encounter in the merger. Note that for one model
including a gaseous disk, we verify the stability in detail in Section
\ref{sec_gasinf}. We will only give a 
short summary of the parameters of our galaxy models and the system of units
used. For details of the method, we refer the reader to \citet{hern1993}. We
assume $G=1$ and the galaxies are intrinsically 
scale-free. As a general rule, all quantities inside this paper are given in
these intrinsically scale free units unless otherwise noted. The conversion to
physical units is performed by scaling a galaxy to some system with known
physical properties. If the progenitors are scaled to the Milky Way, the 
unit time is $13.1$ Myr, unit length is $3.5$ kpc, unit velocity $262$ km
s$^{-1}$ and unit mass is $5.6 \times 10^{10}$ M$_\odot$. 

\begin{table*}
\begin{minipage}{126mm}
\begin{tabular}{|c||r|r|r|r|r|r@{.}l|r@{.}l|}
\hline
\rule[-3mm]{0mm}{8mm}\tb{Model} & \multicolumn{1}{c|}{$\m{N_\mr{\m{disc}}}$} &
\multicolumn{1}{c|}{$\m{N_\mr{\m{bulge}}}$} &
\multicolumn{1}{c|}{$\m{N_\mr{\m{halo}}}$} &
\multicolumn{1}{c|}{$\m{N_\mr{\m{gal}}}$}&
\multicolumn{1}{c|}{$\m{N_\mr{\m{total}}}$} &
\multicolumn{2}{c|}{$\m{h_\mr{\m{star}}}$} &
\multicolumn{2}{c|}{$\m{h_\mr{\m{halo}}}$}
\\\hline\hline
A &   80k &  16k &  160k & 256k &  288k &
0 & 05    & 0 & 071\\\hline
A2 &   80k &  16k &  640k & 736k &  768k &
0 & 05    & 0 & 035\\\hline
B  &  160k &  32k &  320k & 512k&  544k &
0 & 039 & 0 & 056\\\hline
C  &  320k &  64k &  640k & 1024k & 1056k &
0 & 031  &  0 & 044\\\hline
D &  640k & 128k & 1.280k & 2048k & 2080k &
0 & 025   & 0 & 035\\\hline
D2 &  640k  & 128k &  160k & 928k & 960k &
0 & 025   &  0 & 056\\\hline
E  & 1280k & 256k & 2.560k & 4096k & 4128k &
0 & 019 &  0 & 028 \\\hline
\end{tabular}
\caption{Particle numbers of the various components of the pure
  $N$-body simulations. The columns are (from left to right): model
  name, number of disc particles, number of bulge particles, number of
  halo particles, number of particles inside the galaxy which develops
  the tidal tails, total number of particles inside the
  simulation, gravitational softening lengths of disc, bulge and halo,
  respectively. Numbers have 
  been rounded to three digits behind the decimal point.}
\label{tab_res}
\end{minipage}
\end{table*}

The stellar disc has a mass of $m_d=1$. The discs surface density
profile is an exponential with scale length $r_d=1$ and vertical scale height
$z_0=0.2$. The cut off radius of the disc is $r_{max} = 15 r_d$ and
the 
vertical cut off is at $z_{max}=10 z_0$. The Toomre parameter
\citep{toomre1964} at the solar radius is $Q=1.5$. A stellar bulge
component with mass $m_b= 0.2 m_d$ and scale length 
$r_b=0.2r_d$ is used. The cut off radius for the bulge component is
$r_{max,b}=15 r_d$. The dark matter halo is realized as a pseudo isothermal
sphere with mass $m_h=5.8$. It has a core radius of $r_{c,h}=1$ and a
cut off radius of $r_{cut,h}=10$.

From the large sample of $112$ merger simulations of \citet{naab2003} with
different mass ratios and encounter parameters, we chose the encounter which
produces the most pronounced tidal tail and hence is a most favorable case for
TDG formation. The initial conditions used by \citet{naab2003} were consistent
with initial conditions derived from cosmological N-body simulations
\citep[see][]{khochfar2006}. In the encounter we selected, the disc of the
galaxy which produces the tidal tails is 
aligned with the orbital plane of the encounter, the galaxies have equal mass
and the encounter is prograde. The galaxies are put on a parabolic orbit, the
initial separation between their centers of mass is 30. The pericenter
distance of the encounter is $r_p=2$. Together with their masses, this
determines the orbit completely.

In order to study the formation and subsequent evolution of tidal
tails in detail, the resolution of the progenitor galaxy from which
the tidal tail emerges should be maximized as far as possible in a
simulation. The second galaxy, on the other hand, only acts as a
perturber. It creates the tidal forces which strip the future tail
material from the progenitor disc. The internal structure, dynamics and
orientation of the perturbing galaxy are 
irrelevant for the tidal tail as long as the tidal forces during the
encounter are reproduced accurately enough. In order to be able to maximize
the resolution of the progenitor galaxy, we therefore only resolve the
perturbing galaxy comparatively coarse. This approach allows the study
of only one tidal arm per merger, but this in greater detail, which is
the aim of this work. We have confirmed the validity
of this approach in a test simulation in which both galaxies were
resolved at the same level. The matter which is found inside the
tidal arm of one galaxy contains no particles of the perturbing galaxy. The
tidal arm forms inside halo material of the galaxy to which also the stellar
material of the tail once belonged.

We set up the perturbing galaxy with a 
resolution of 10000 particles for the stellar disc, 2000 particles for
the bulge component and 20000 for the dark matter halo. Both the
perturbing galaxy as well as the one which develops the tidal tail
share the parameters for the mass and extent of the components as
described above. However, the galaxy which develops the tidal tails is
realized at much higher resolution than the perturbing one. The
particle numbers for our different models are summarized in Table
\ref{tab_res}. The lowest resolution model, A, has already more
particles inside one galaxy than the entire simulation of the
classical \citet{barnes_hern1992} paper and the follow up simulations
of \citet{barnes_h1996}. As far as we know, our highest resolution simulation,
model E, is the highest resolution simulation of a tidal tail published so
far.

\subsection{Numerical Method}
\label{sec_params}

The simulations were performed with our hybrid N-body / SPH code
VINE. The code will be described in detail in a forthcoming paper (Wetzstein
et al., in preparation), so here we will only give a short description of the
numerical algorithms used. VINE uses a binary, mutual nearest neighbor tree
for the efficient calculation of gravitational forces
\citep{appel85,press86}. A distant 
tree node $j$ is accepted for the interaction list of particle $i$ if their
distance $r_{ij}$ satisfies the condition $r_{ij} > h_i + h_j/\theta$ where
$h_i$ and $h_j$ are the physical size of the particle and node, 
respectively. For the tolerance parameter $\theta$, we adopted a value of
$0.7$ for our simulations presented here. The time integration was carried out
using a leapfrog integrator in combination with an individual time step scheme
which allows the 
code to follow very high density contrasts in the simulated system very
efficiently. The individual time step scheme is similar to the blockstep
scheme of the TREESPH code \citep{hern_katz89}. The timestep $t_{n+1}$
of a particle is computed using the minimum of three different criteria:
$t_{n+1}=\min(\tau_1 \sqrt{h/|\m{a}|}, \tau_2 h/|\m{v}|, \tau_2
|\m{v}|/|\m{a}|)$ where $h$ is the gravitational softening length, $\m{v}$ the
velocity and $\m{a}$ the acceleration of the particle. The $\tau_i$ are
tolerance parameters, for which we used $\tau_2=0.14$ and $\tau_2=1$. Our
highest 
resolution simulations required time steps being a factor of more than
$32000$ smaller than the maximum timestep in the scheme. The SPH scheme in
VINE follows the SPH formulation of \citet{benz90}. The artificial viscosity
included the additional terms suggested by \citet{balsara90, balsara95} to
reduce entropy generation in shear flows. Our choice of parameters of the code 
allowed us to keep the error in the total energy of the system lower than
$0.3\%$ in our $N$-body simulations. 

VINE is fully parallelized with OpenMP. The gravitational force calculation,
which is the most time consuming part of a simulation, scales linear up to
$100$ processors and the second most time consuming part, i.e. the SPH
calculation scales equally well.
The simulations have been run in parallel on 32 processors of our SGI Altix
supercomputer at the University-Observatory, Munich, as well as the IBM Regatta
supercomputer at the Rechenzentrum of the Max Planck Society, Garching. 

\subsection{Gravitational Softening}
\label{sec_soft_setup}

As we want to study the collapse of objects inside the tidal arms, the
particle noise is an important factor
\citep{barnes_hern1992}. An additional effect here is the
gravitational force softening used. The optimal choice for the force
softening, regarding its functional form as well as the particular
softening scale used as a function of particle number have been a
subject of intensive debate in the literature \citep[see e.g.][ and
  references therein]{merritt96,romeo97,afl2000,dehnen2001}. We 
use the SPH spline kernel to soften the gravitational forces of all
particles. It has the advantage that outside the softening length it exactly
resembles the Newtonian potential. The benefits of this choice
over Plummer softening \citep{aarseth1963} have been discussed by
\citet{dehnen2001}.

Regarding the choice of the softening lengths,
there exist no studies particularly dedicated to mergers of interacting
galaxies. We have thus used empirical values for the softening of the
stellar disc in model A. For a merger with a disc of 80000 particles, our
choice for the softening length is $h=0.05$ \citep[see e.g.][]{naab2003},
which guarantees a stable thin disc after several rotation periods. In
order to obtain 
softening lengths for the bulge and halo component of the 
galaxy, we have adopted the point of view of force softening being
equivalent to giving a point mass a finite mass
distribution. Particles with higher mass should thus have increased
softening lengths as well. This idea was applied to every component
(and thus particle species with a given mass) of the galaxy by scaling
the smoothing length of $h_{80000}=0.05$ for our particles in the disc
of model A (80000 particles) with mass 
$m_{80000}=m_d/80000$ to other masses according to $h_{new}=h_{80000}  
(m_{new}/m_{80000})^{(-1/3)}$. The effects of two body relaxation and
associated disc 
heating should hence be decreased, together with similar effects of
dark matter halo particles penetrating the tidal tail at later stages of
the simulation. The corresponding softening lengths used for the
model galaxies are listed in Table \ref{tab_res}. The softening
lengths of the components of the perturber galaxy have been adjusted
in the same way, resulting in a fixed softening length of $h_\mr{disc}=0.1$
for the stellar disc, $h_\mr{bulge}=0.1$ for the bulge and
$h_\mr{halo}=0.1797$ for the dark matter halo.
\begin{figure*}
  \begin{center}
    \leavevmode
    \epsfig{file=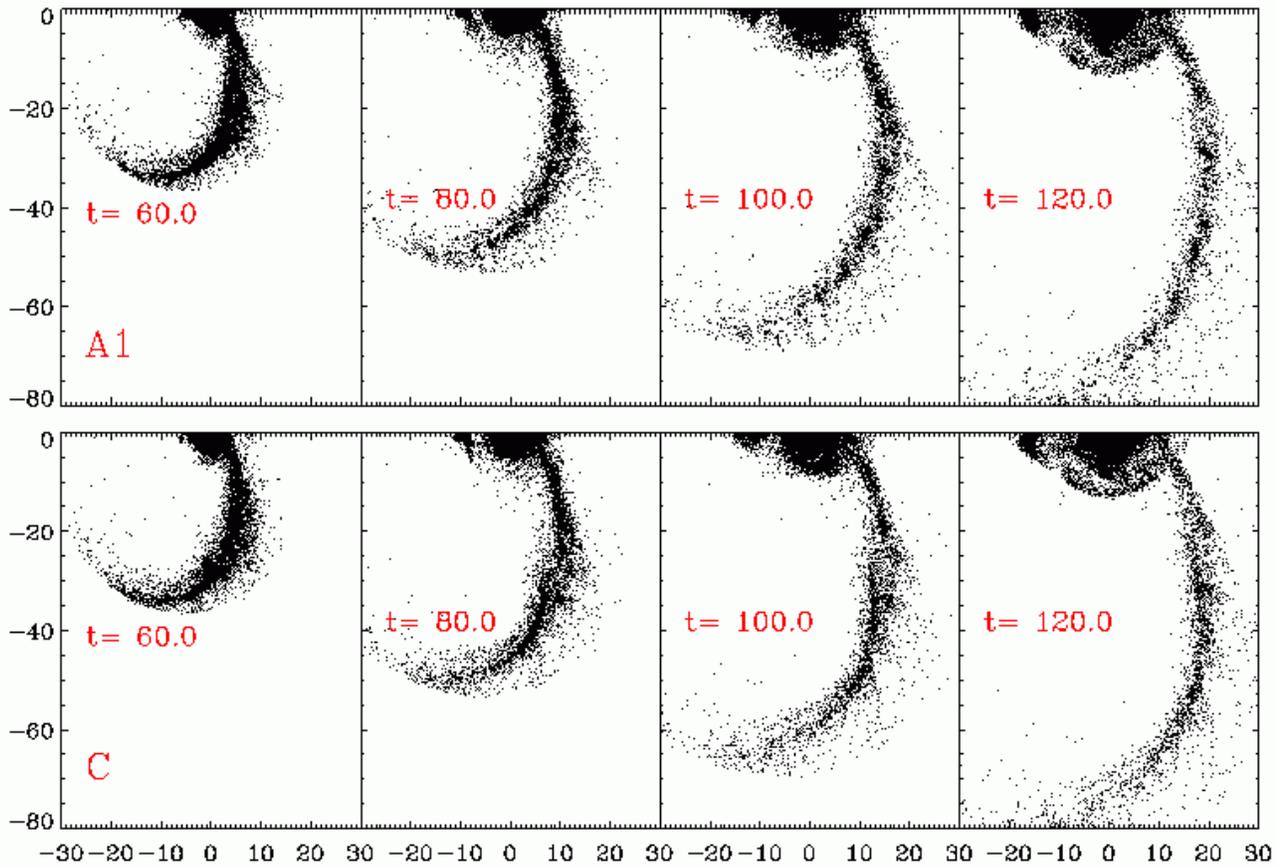,width=17cm,height=12.3cm}
        \caption{Time evolution of stellar material in model A (upper panel),
                 and model C (lower panel), projected on the orbital plane. In
                 model A, collapsed objects form in the tidal arm (last plot
                 in upper panel), in contrast to model C (see last plot
                 in lower panel).}     
   \label{fig_snap80k_320k}
   \end{center}
\end{figure*}

In any resolution study like the one presented here, softening 
can affect the results. This can only be addressed by numerical experiments,
i.e. actually varying the softening lengths \ti{and} the particle numbers. We
have carried out such tests in order to assure the validity of our results
against softening effects, see Section \ref{sec_soft_result} for a detailed
discussion.

\section{RESULTS OF THE RESOLUTION STUDY}
\label{sec_res}

\subsection{Influence of Increased Progenitor Resolution}
\label{sec_totres}

In order to study the effects of numerical resolution, we have conducted a
sample of simulations with different resolutions of the same configuration,
as described in Section \ref{sec_ic}. Starting from a low resolution
simulation with 80000 particles in the disc of the progenitor galaxy
(model A), we increased the mass resolution up to a factor of 16 (model
E). This should allow us to assess the hypothesis of
\citet{barnes_hern1992} that particle noise is one of the sources of the
density perturbations which lead to the gravitational collapse of
objects. 

All the simulations have been run from $t=0$ until $t=200$, at which
the central part of the remnant galaxy has relaxed into a new
equilibrium configuration. The first close encounter of the two galaxies
occurs at $t\approx25$ and the galaxies merge until $t\approx55$.
The time evolution of model A from $t=60$
to $t=120$ is shown in the upper panel of Figure
\ref{fig_snap80k_320k}. In this and the following particle scatter plots, we
show only a randomly chosen subset of 40000 particles to allow for better
viewing. The tidal tail initially goes through a phase
of rapid expansion. Later, its inner parts fall back onto the merger remnant,
while the outer parts still expand and move further away from the remnant.

Several fragments formed within the tidal tail and collapsed
afterwards. They decoupled from the surrounding tail and became
self-gravitating entities. For our analysis, we identify the objects with a
friends-of-friends algorithm and check every object in the
tail for self-gravity. We require the object to reach central stellar densities
similar to those of observed dwarf galaxies, otherwise we don't consider the
object as a possible candidate for a TDG. We adopt a stellar density threshold
of $\rho_0 \ge 0.013$ 
M$_\odot/$pc$^3$, corresponding to $\rho_{0,code} \ge 0.01$ in our code
units. This choice is motivated by the low central density tail of observed
dwarf galaxies \citep{mateo1998}, lowered slightly to yield a conservative
lower limit.
The formation of the objects in our model A is very
similar to those in the simulation of \citet{barnes_hern1992}. We can
confirm their result that dark matter does not significantly
contribute to the overall mass of those objects and find a fraction of
$\le 3\%$.

Model B evolves very similar to model A, hence we don't show its
evolution in a separate figure. Several collapsed objects form inside
the tidal tails, but their distribution along the tidal tail is different from
the one in model A. This indicates that results of
simulations on TDG formation are affected by numerical resolution, as
one would expect if particle noise is affecting the gravitational
collapse.

\begin{figure}
  \begin{center}
    \leavevmode
    \epsfig{file=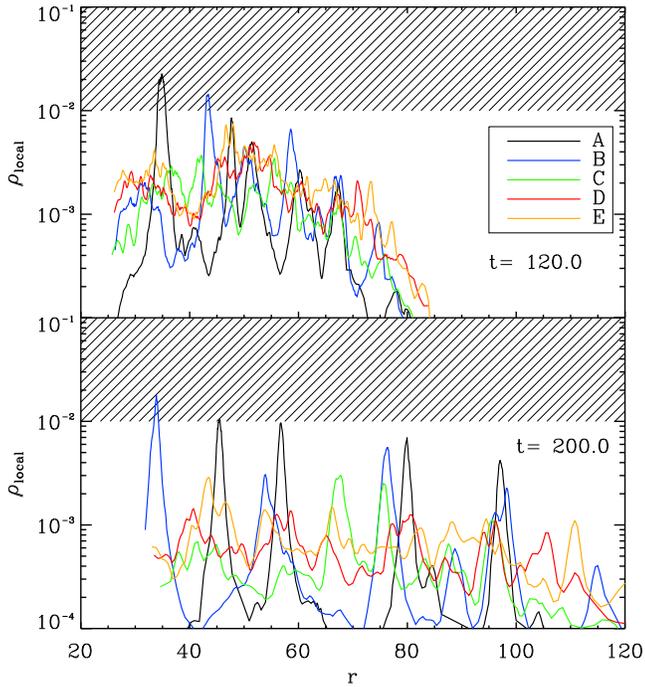,width=8.5cm}
        \caption
               {Local stellar density $\rho_{local}$ in the tidal tail
               vs. distance $r$ from the center of the remnant at 
               t=120 and t=200 for models A, B, C, D, and E. The
               shaded region at the top of both panels indicate a lower limit
               on central stellar densities of dwarf galaxies, see text for
               more details.}      
   \label{fig_rho_normsoft}
   \end{center}
\end{figure}

Increasing the mass resolution by another factor of two seems to
suppress the gravitational collapse, as can be seen from the lower
panel of Figure \ref{fig_snap80k_320k} which shows the time evolution
of model C. Apparently some regions in the tidal tail are more dense
than others, but no collapsed objects as in the lower resolution
models are formed. In order to study this result more quantitatively, Figure
\ref{fig_rho_normsoft} shows the density inside the tidal arm 
as a function of distance from the center of the merger remnant for models A,
B, C, D, and E. The upper panel shows the density at $t=120$, the latest time
shown in Figure \ref{fig_snap80k_320k}. In addition, the lower panel shows the
densities along the tidal tail at $t=200$, a rather late phase of the tail
evolution which allows us to assess the final stage of collapse along the
tidal arm. 
We obtain the density in the tidal arm from the stellar components alone, as
the contribution of dark matter to the total mass of the tidal arm is
negligible. For every 
star particle, we compute its distance from the center of the merger
remnant. The density of the particle is computed in a way similar to
the SPH density calculation: For every particle, we search for the 50
nearest neighbouring particles and compute a mean density of the
sphere containing those particles. For the plots in Figure
\ref{fig_rho_normsoft}, we create bins of equal mass in order to allow for a
reasonable  comparison between the different resolutions. Every point
in the profiles of Figure \ref{fig_rho_normsoft} represents the
maximum stellar density of the corresponding bin. The fixed mass 
inside each bin is chosen as $1.25 \times 10^{-4}$, corresponding to
10 particles in model A and 160 in model E, respectively.

Figure \ref{fig_rho_normsoft} shows that with increased resolution, no
collapsed objects form inside the tail anymore. Note
that the innermost objects in the tail at $t=120$ have fallen back into the
remnant galaxy at $t=200$ and hence are no longer present in the lower panel
of Figure \ref{fig_rho_normsoft}. The tidal tail in model B, which also forms
collapsed objects, exhibits a similar evolution as model A. However, the
growth of overdensities is delayed with respect to model A. 

As was already indicated in Figure \ref{fig_snap80k_320k}, the tidal tail of
model C does produce overdense regions, but these hardly develop into
collapsed objects. Following the trend of model B, the growth of
overdensities is delayed further. Only at very late stages, some of these
overdensities developed a high density contrast relative to their
surroundings. 

Models D and E follow this trend as well. The mass
assembly of overdense regions is delayed even further, and the
associated regions do not gravitationally collapse to form distinct
entities, such as the objects in model A or B. Comparing the density
evolution of the different models further, there are remarkable
differences between models A to D. However, the density of the tidal
tails in model D and E do not differ much and agree considerably well
especially for the early stages of the tail evolution. We interpret
this as a sign that for the time scale until $t=200$ studied
here, the simulations start to converge somewhere around
the resolution of model E.

\begin{figure}
  \begin{center}
    \leavevmode
    \epsfig{file=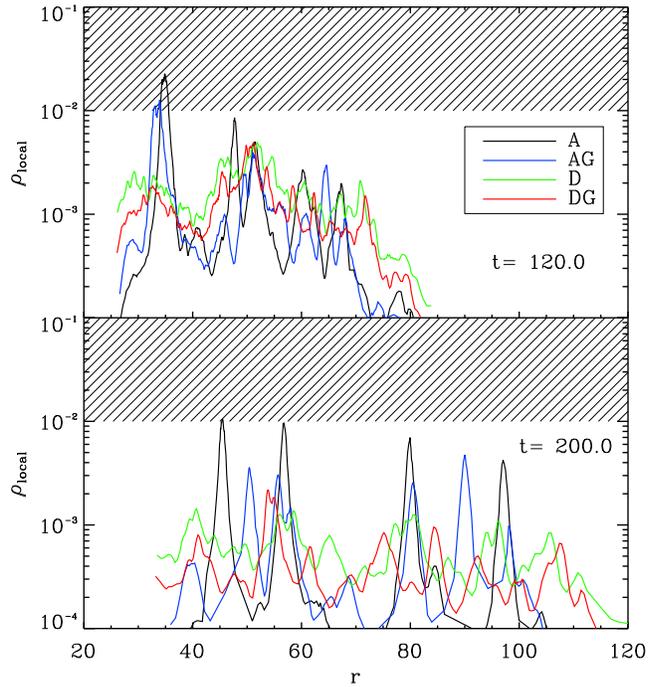,width=8.5cm}
        \caption
               {Same as Figure \ref{fig_rho_normsoft} for models A and D as
               well as the additional models AG and DG, which were set up with
               a different random seed and simulated with Gadget2.}
   \label{fig_rho_gad}
   \end{center}
\end{figure}
In order to assure that our result is independent of the simulation code used,
we re-simulated models A and D, using a different random seed for the setup and
the Gadget2 code \citep{springel_g2} for the time evolution. We used an
integration accuracy parameter of $0.02$ and a tolerance parameter for the
tree accuracy of $10^{-4}$. The results of the Gadget2 run are
shown in Figure \ref{fig_rho_gad}. Again we see that the density in the peaks
decreases when resolution is increased.

We thus conclude from this analysis that the gravitational
collapse of objects inside a tidal arm, which has been reported for pure
$N$-body simulations \citep{barnes_hern1992,barnes_h1996}, is strongly
suppressed if the simulations have high enough resolution. The TDG candidate
objects formed in models A and B are artifacts of the 
poor resolution in those models. Furthermore, only these two models are able to
form TDG candidate objects with central densities similar to those of observed
dwarf galaxies \citep{mateo1998}. All other models shown in Figure
\ref{fig_rho_normsoft} fail 
to reach such central densities in objects in the tidal arm. As we described
in Section \ref{sec_ic}, the particular encounter which we study here is a
most favorable case for the formation of extended tidal arms. Hence it is
unlikely that other encounter parameters for the merger would work
better to produce TDGs. We can conclude that
pure $N$-body simulations of merging galaxies fail to generate collapsed
objects in the tidal tails which resemble dwarf galaxies, provided that the
resolution of the simulations is high enough.

\subsection{Effects of Gravitational Softening on Tidal Tail}
\label{sec_soft_result}

In resolution studies like the one presented here the total number
of particles is not the only parameter affecting the local gravitational
potential. The other key parameter in this context is the gravitational
softening length. Both, the softening length and the number of particles can
cause two body relaxation effects such as artificial heating of an N-body
system. For example, a system realized with few particles can have a very noisy
gravitational potential if the softening lengths are small, or a rather smooth
potential if the softening length is increased considerably. Especially if the
gravitational collapse of small structures is to be studied, both softening
length and number of particles should ideally be explored to evaluate their
effects on the results obtained from such simulations. As the exploration of
this two parameter space is usually very expensive in terms of cpu time, such
studies are rather rare. The usual approach is to resolve the system at hand
with the highest particle number affordable under given cpu time constraints
and to simulate this setup with a hopefully well chosen gravitational
softening length. The latter is extremely dependent on the simulator's personal
experience, as systematic studies of force softening as a function of particle
number are rare and usually only exist for certain well behaved mass
distributions which in turn are often very different from the system to be
studied \citep[see e.g.][and references
therein]{merritt96,romeo97,theis1998,afl2000,dehnen2001}. 

Nevertheless, we have explored the effects of different gravitational
softening lengths for some of our models. The set of simulations used for this
purpose is summarized in Table \ref{tab_soft}.
\begin{table}
\begin{center}
\begin{tabular}{|l|r@{.}l|r@{.}l|}
\hline
\rule[-3mm]{0mm}{8mm}\tb{Model} & \multicolumn{2}{c|}{$\m{h_\mr{\m{star}}}$} &
\multicolumn{2}{c|}{$\m{h_\mr{\m{halo}}}$} 
\\\hline\hline
C    & 0 & 031  & 0 & 044  \\\hline
CL4  & 0 & 0078 & 0 & 011  \\\hline
CH4  & 0 & 124  & 0 & 176  \\\hline
D    & 0 & 025  & 0 & 035  \\\hline
DL4  & 0 & 0063 & 0 & 0088  \\\hline
DL16 & 0 & 0016 & 0 & 0022  \\\hline
\end{tabular}
\caption{Gravitational softening lengths of the star and dark matter particles
  in simulations used to study softening effects. Apart from different
  softening lengths, models CL4 and CH4 are identical to model C, DL4 and DL16
  are otherwise identical to model D. In CL4, the softening lengths have been
  decreased by a factor of 4 compared to model C, while in CH4 they have been
  increased correspondingly. Model DL4 has a factor of 4 lower softening than
  D, DL16 has a factor of 16 lower softening than D.}
\label{tab_soft}
\end{center}
\end{table}

We have repeated the simulation of model C, but with a factor of 4 lower
(model CL4) and a factor of 4 higher (model CH4) softening length. A similar
set of models has been created for case D, with factors of 4 and
16 lower softening lengths in models DL4 and DL16, respectively. Apart from
the different softening lengths, all other parameters of those models have been
inherited from models C and D, respectively.
\begin{figure}
  \begin{center}
    \leavevmode
    \epsfig{file=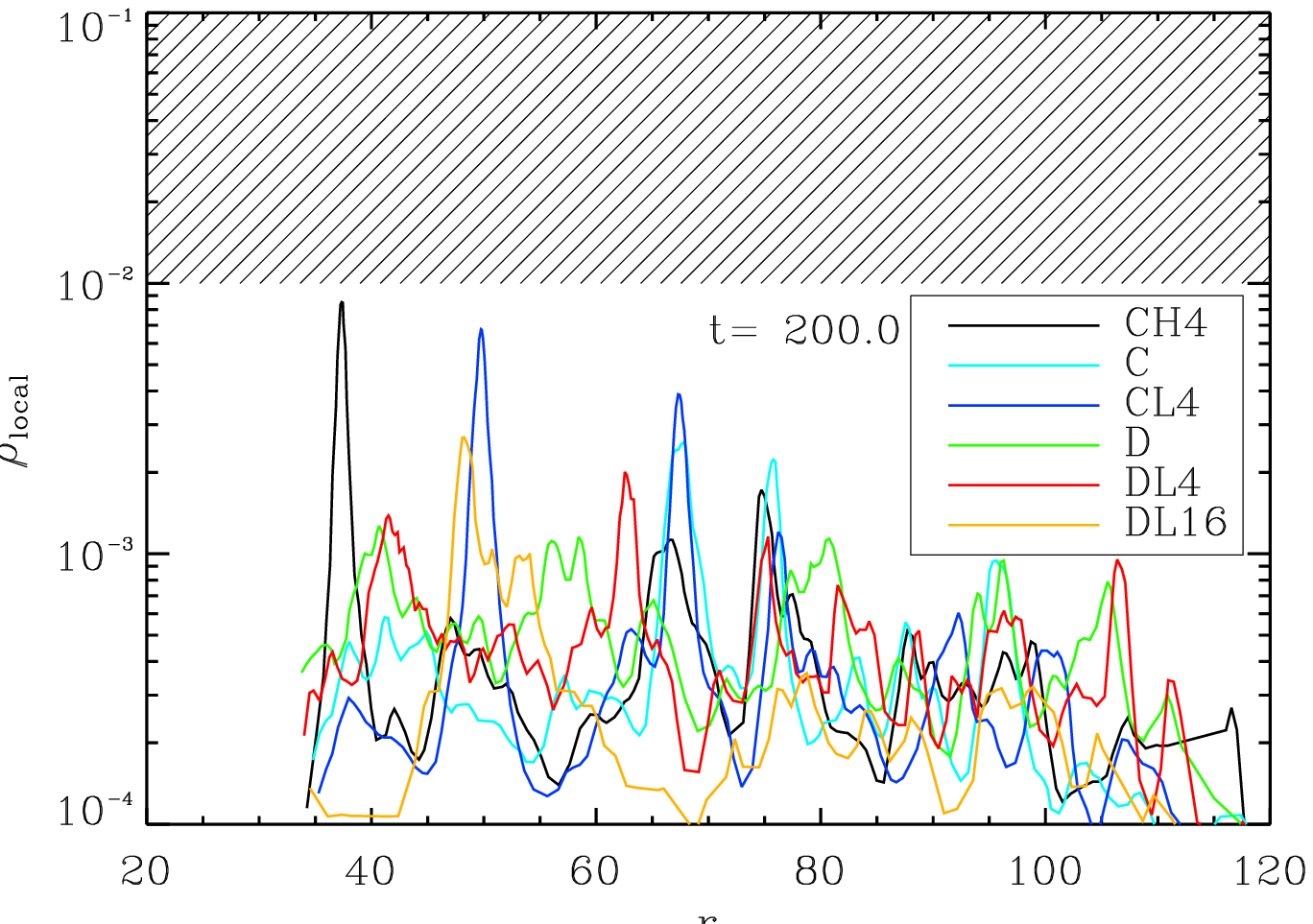,width=8.5cm}
        \caption
               {Same as Figure \ref{fig_rho_normsoft} for models CH4, C, CL4,
               D, DL4 and DL16 at time $t=200$.}
   \label{fig_rho_lowsoft}
   \end{center}
\end{figure}
Figure \ref{fig_rho_lowsoft} shows the density in the tidal tails of models C,
CH4, CL4, D, DL4 and DL16 as a function of distance from the center of the
merger remnant. 

Although different softening lengths lead to the formation of different high
density regions in the tidal arm, none of these regions reach densities of
dwarf galaxies. We can therefore conclude that our particular choice
of softening lengths in models A to E has not artificially suppressed the
formation of bound objects with central densities similar to dwarf galaxies.

\subsection{Influence of Progenitor Dark Matter Halo}

From the analysis presented so far, it is not clear which component of the
progenitor galaxy, i.e. disc, bulge or halo, plays the key role 
in triggering the collapse of objects in the tidal arm at low resolutions. The
bulge component is 
very unlikely to play an important role here, as the vast majority of the
bulge particles is located inside the merger remnant rather than the tidal
tail. So we need to address the question if the resolution of the stellar disc
of the progenitor galaxy is important or if the resolution of the dark matter
halo of the progenitor galaxy plays a crucial role in this
respect. The halo component  could cause perturbations in the stellar
material of the tidal arm which then could grow and finally collapse. However,
it could as well be the resolution of the stellar component itself which
causes the collapse.

In order to study the effects of both, we have created two additional models,
A2 and D2. They are identical in setup to model A and D, respectively, but
their dark matter halos have been realized with different particle numbers
(see Table \ref{tab_res}). In model A2, the halo bas been sampled with 640000
particles, while in model D2 the halo has been sampled with only
160000. Model A2 shares the resolution of the stellar disc with
model A, but the dark halo is resolved with four times more particles
than in A, thereby giving A2 the halo resolution of model C. Model
D2 also shares the resolution of the stellar disc with model D,
but its dark matter halo is resolved with eight times less particles,
giving it the halo resolution of model A.

\begin{figure}
  \begin{center}
    \leavevmode
    \epsfig{file=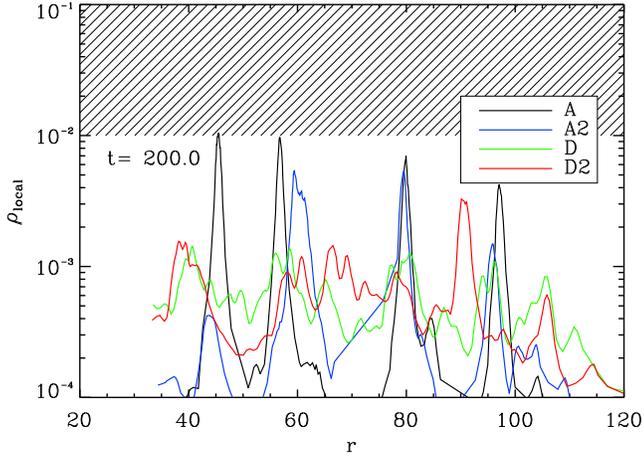,width=8.5cm}
        \caption{Same as Figure \ref{fig_rho_normsoft} for models A, A2, D and
    D2 at $t=200$.}
   \label{fig_rho_disc_halo}
   \end{center}
\end{figure}

Apart from the different disc / halo resolutions discussed above, the
simulations of models A2 and D2 shared the identical set of parameters
as used in the other models. The density profiles of the tidal tails for
models A2 and D2 are compared with the results of models A and D in Figure
\ref{fig_rho_disc_halo}.  

In the tidal tail of model A2, the density evolution of
the outer tail is very similar to that of model A. In the inner
region of the tail, however, instead of two overdensities in model A
there forms only one overdensity in model A2. The peak densities are
almost identical. The overall behavior is similar in the sense that
dense objects are actually formed in both A and A2. 

Model D2 also evolves rather similar to D. There are only small
differences between the two simulations. In model D2 there forms a
density peak at $r \approx 91$ which is not present in
D. Apparently, the formation of this overdensity was caused by the lower
resolution dark matter halo of D2. But apart from this one region, the
results of D and D2 are similar. Collapse of objects along the tidal
tail is still highly suppressed.

If shot noise from the dark halo plays an important role regarding
the collapse of objects in simulated tidal tails, this should be visible in
model A2 as well as D2. The tidal tail in D2 results from a stellar disc which
is as well resolved as in model D, which did not form collapsing objects in
its tidal tail. Thus if in model D2 collapsed objects would have formed in the
tail, this would indicate that shot noise of the dark halo plays an important
role. In addition, as model A2 is otherwise identical to model A, its halo has
been given the resolution of model C. So if the
resolution of the halo would dominate the collapse of objects in the tidal
tail, then this collapse should be effectively suppressed in model
A2. However, model A2 still shows  collapsed objects as in model A while at
the same time the lower resolution halo of model D2 only triggers the
collapse of one region in the tidal tail. The rest of the tail is very
similar to model D.

In summary, the high resolution dark matter halo in model A2
did not at all suppress the collapse of structures inside the tidal
tail. The evolution was very similar to the model with low resolution
dark halo, A. On the other hand, the high resolution stellar disc of
model D2 did not exhibit collapsing objects like in model A, for example,
although the disc in D2 resided in a low resolution dark matter halo. The
evolution of the tidal tail was comparable to that of model D, apart from one
additional overdense region which formed in D2.

If we would have chosen a single softening length for all particles in the
simulation, as is often done when using Plummer softening, this
softening length would need to be large enough to keep the most massive
particles, i.e. those in the halo, collisionless. This choice would probably
result in larger softening lengths than those which we used, possibly
suppressing local collapse to some extent. 
\begin{figure}
  \begin{center}
    \leavevmode
    \epsfig{file=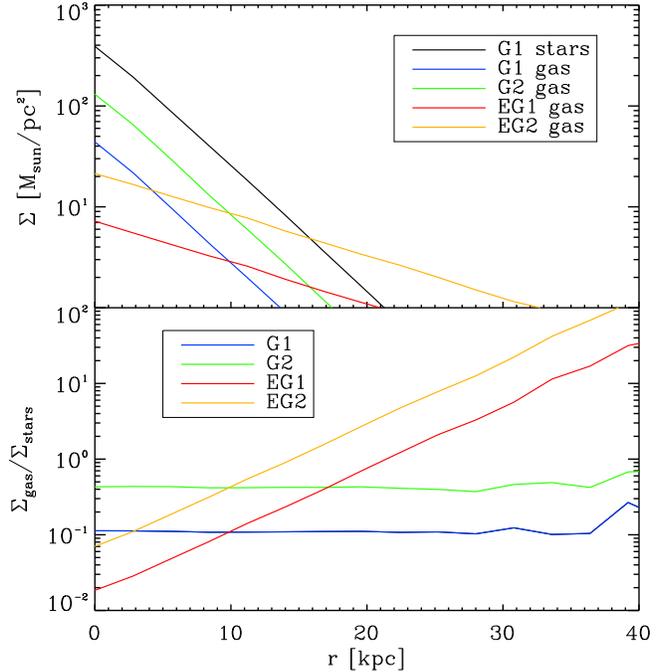,width=8.5cm}
        \caption{Surface density distribution of gas and stars in the
    progenitor galaxies for models G1, G2, EG1 and EG2 (upper panel) and ratio
    of stellar to gas surface density in  those models (lower panel). The
    region with $r \ge 30$ kpc is affected by low particle numbers in the
    corresponding radial bins. Note that for ease of comparison, the model has
    been scaled to the Milky Way and quantities are given in physical
    units.}
   \label{fig_gas_surf}
   \end{center}
\end{figure}
\begin{figure}
  \begin{center}
    \leavevmode
    \epsfig{file=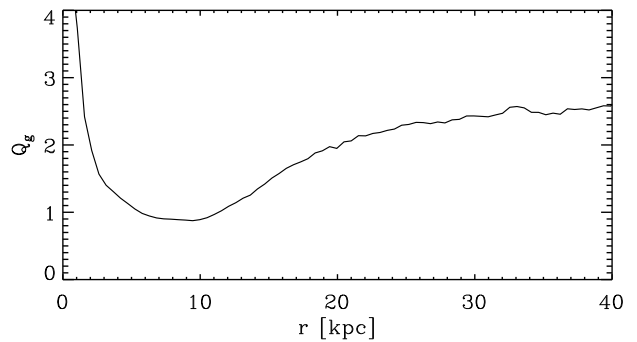,width=8.5cm}
        \caption{Initial radial profile of the Q parameter for the gas
    disk. Note that for ease of comparison, the model has been scaled to the
    Milky Way and quantities are given in physical units.}     
   \label{fig_qprof}
   \end{center}
\end{figure}

We conclude that the major effect leading to artificial numerical
collapse of structures inside tidal tails of pure $N$-body simulations
is the resolution of the stellar disc of the progenitor galaxy which
forms the tidal tail. With high enough resolution, collapse of
such structures is strongly suppressed. The resolution of the dark halo has
some influence on the formation of overdense regions, but the effect is
small compared to that of the stellar disc resolution.


\subsection{Gas Dynamical Effects on the Formation of TDGs}
\label{sec_gasinf}


\begin{table}
\begin{center}
\begin{tabular}{|c||r|r@{.}l|r@{.}l|}
\hline
\rule[-3mm]{0mm}{8mm}\tb{Model} & \multicolumn{1}{c|}{$\m{N_\mr{\m{gas}}}$} &
\multicolumn{2}{c|}{$\m{M_\mr{\m{gas}}}$} &
\multicolumn{2}{c|}{$\m{r_\mr{\m{gas}}/r_\mr{\m{star}}}$}
\\\hline\hline
G1  & 32k & 0 & 1 & 1 & 0  \\\hline
G2  & 45k & 0 & 3 & 1 & 0  \\\hline
EG1  & 32k & 0 & 1 & 3 & 0  \\\hline
EG2  & 45k & 0 & 3 & 3 & 0  \\\hline
\end{tabular}
\caption{Parameters of the gas simulations. The first column gives the
  number of SPH particles used, the second the total mass of the gas
  component and the third the scale length of the gas disc in units of
  the stellar scale length.} 
\label{tab_gas}
\end{center}
\end{table}

The resolution study of pure $N$-body simulations presented in Section
\ref{sec_totres} resulted in the interesting fact that in
such simulations, tidal dwarf galaxies are suppressed if the
resolution of the simulations is high enough. 
\begin{figure}
  \begin{center}
    \leavevmode
    \epsfig{file=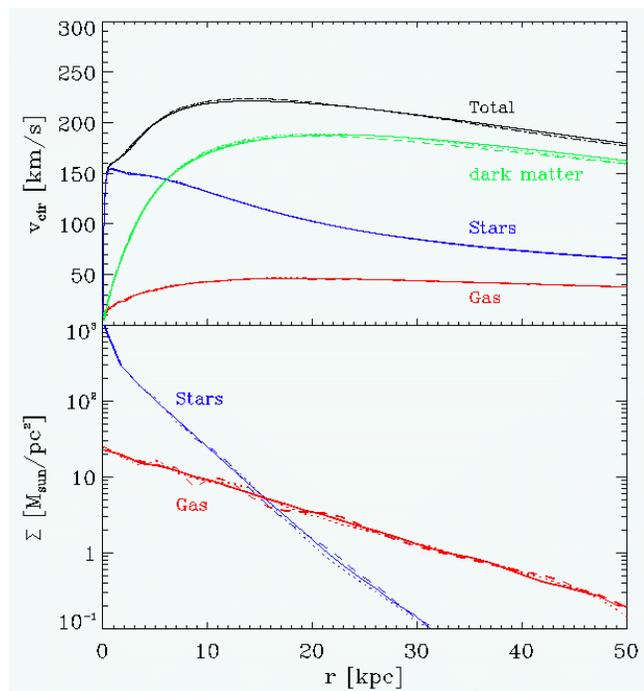,width=8.5cm}
        \caption{Circular velocity (upper panel) and surface density (lower
    panel) for the high resolution galaxy in model EG2 when evolved in
    isolation. Note that for ease of comparison, the model has been scaled to
    the Milky Way and quantities are given in physical units. The
    solid lines show the initial state, the dotted lines are at 
    $t=327.5$ Myr and the dashed lines at $t=589.5$ Myr.}     
   \label{fig_stability}
   \end{center}
\end{figure}

\begin{figure}
  \begin{center}
    \leavevmode
    \epsfig{file=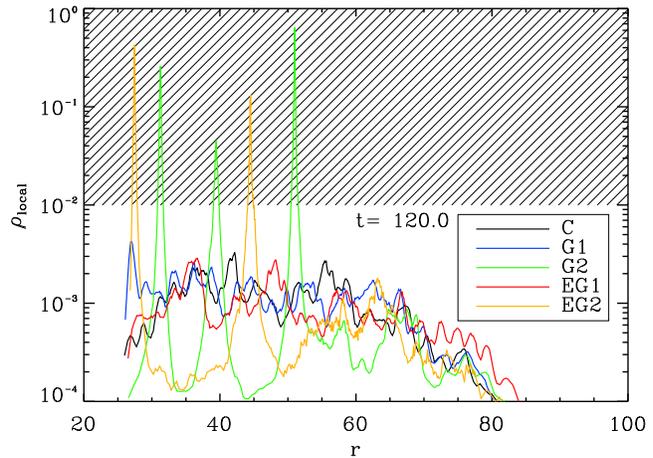,width=8.5cm}
        \caption
               {Same as Figure \ref{fig_rho_normsoft} for models C, G1, G2, EG1
               and EG2 at $t=120$.}   
   \label{fig_rho_gas}
   \end{center}
\end{figure}
As gas is observed in practically any tidal tail (see Section
\ref{tdg_obs}), it could possibly be the lacking ingredient in
simulations of tidal dwarf galaxy formation. \citet{barnes_h1996}
argue that gas would not play an important role in triggering the
collapse of structures inside a tidal tail. However, their
argumentation was based on their result that pure $N$-body simulations
of tidal tails, without any gas, all by themselves already lead to the
formation of collapsed objects and hence gas would not be required for
this process. As we have shown that for our models,
such collapse in pure $N$-body
simulations is artificial and induced by too low resolution, it
seems viable that we explore in how far gas could actually be important
for the collapse of structures in tidal tails.

In order to study the effects of gas on tidal tails, we take the setup
of model C (see Table \ref{tab_res} for the details) and add a gaseous disc
component using the SPH 
approach. Model C was the lowest resolution simulation in our sample
which did not show collapsed objects like those in models A and B in
the tail. In addition, including 
hydrodynamic effects in a merger simulation significantly increases
the amount of CPU time required for completion. So model C is a good
starting point for the study of gas dynamical effects when taking the
increased computational expense into account.
\begin{figure*}
  \begin{center}
    \leavevmode
    \epsfig{file=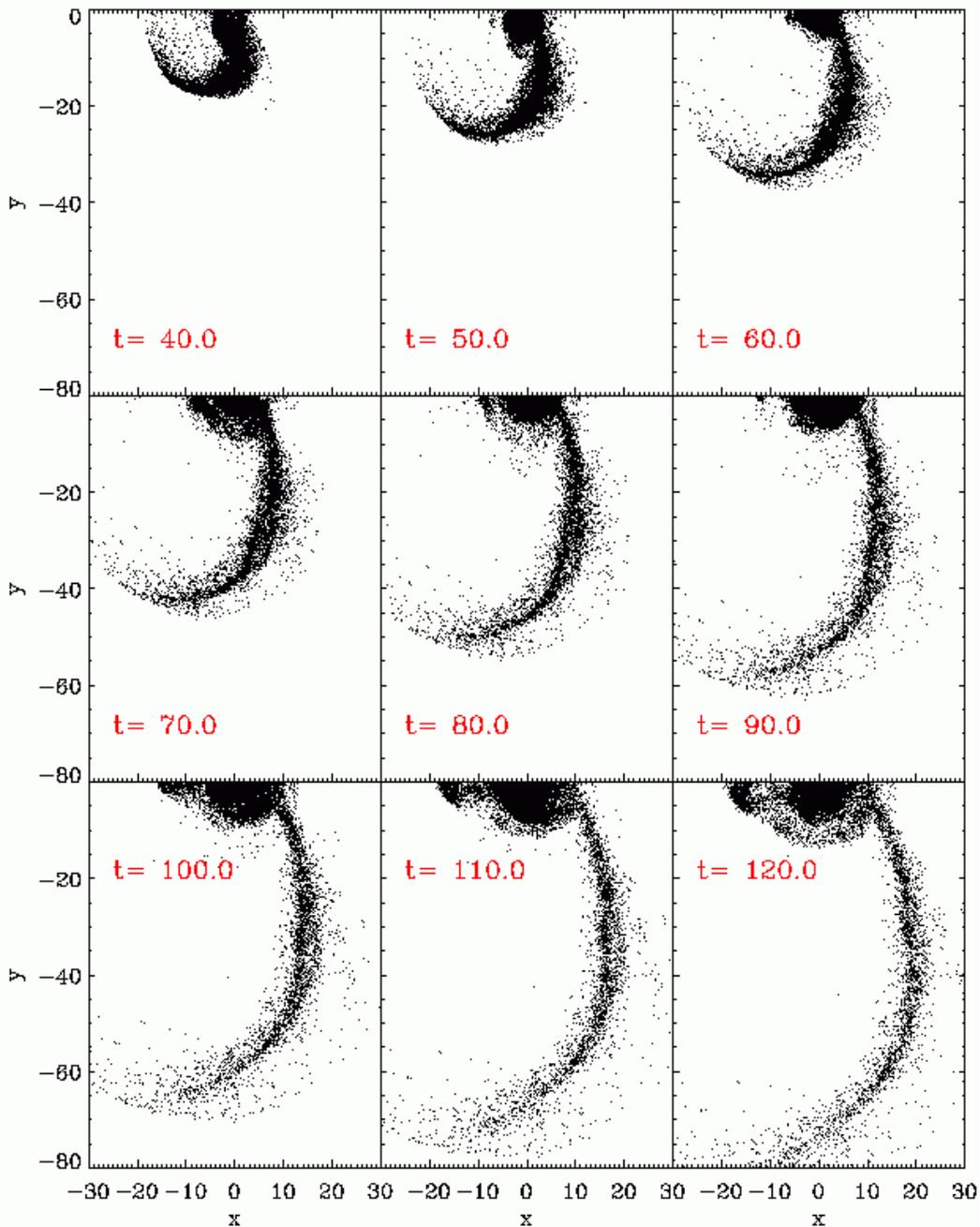,width=14cm}
        \caption{Time evolution of stellar component in model G1, projected on
               the orbital plane. No collapsed objects form.}
   \label{fig_11G320KQHG1STH}
   \end{center}
\end{figure*}
\begin{figure*}
  \begin{center}
    \leavevmode
    \epsfig{file=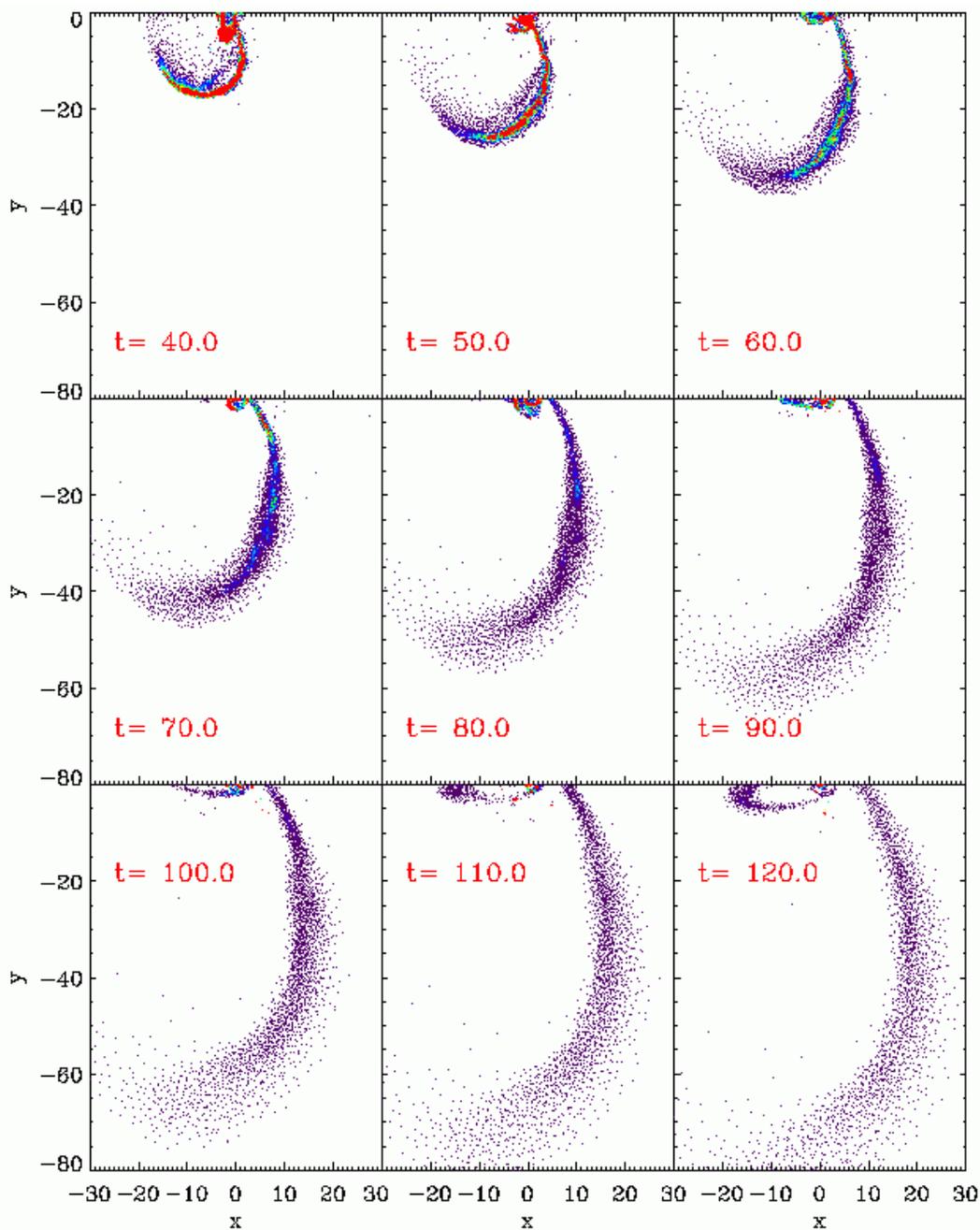,width=14cm}
        \caption{Time evolution of gas in model G1, projected on
               the orbital plane. The logarithmic gas densities are color
               coded. No collapsed objects form.} 
   \label{fig_11G320KQHG1STH_gas}
   \end{center}
\end{figure*}

\begin{figure*}
  \begin{center}
    \leavevmode
    \epsfig{file=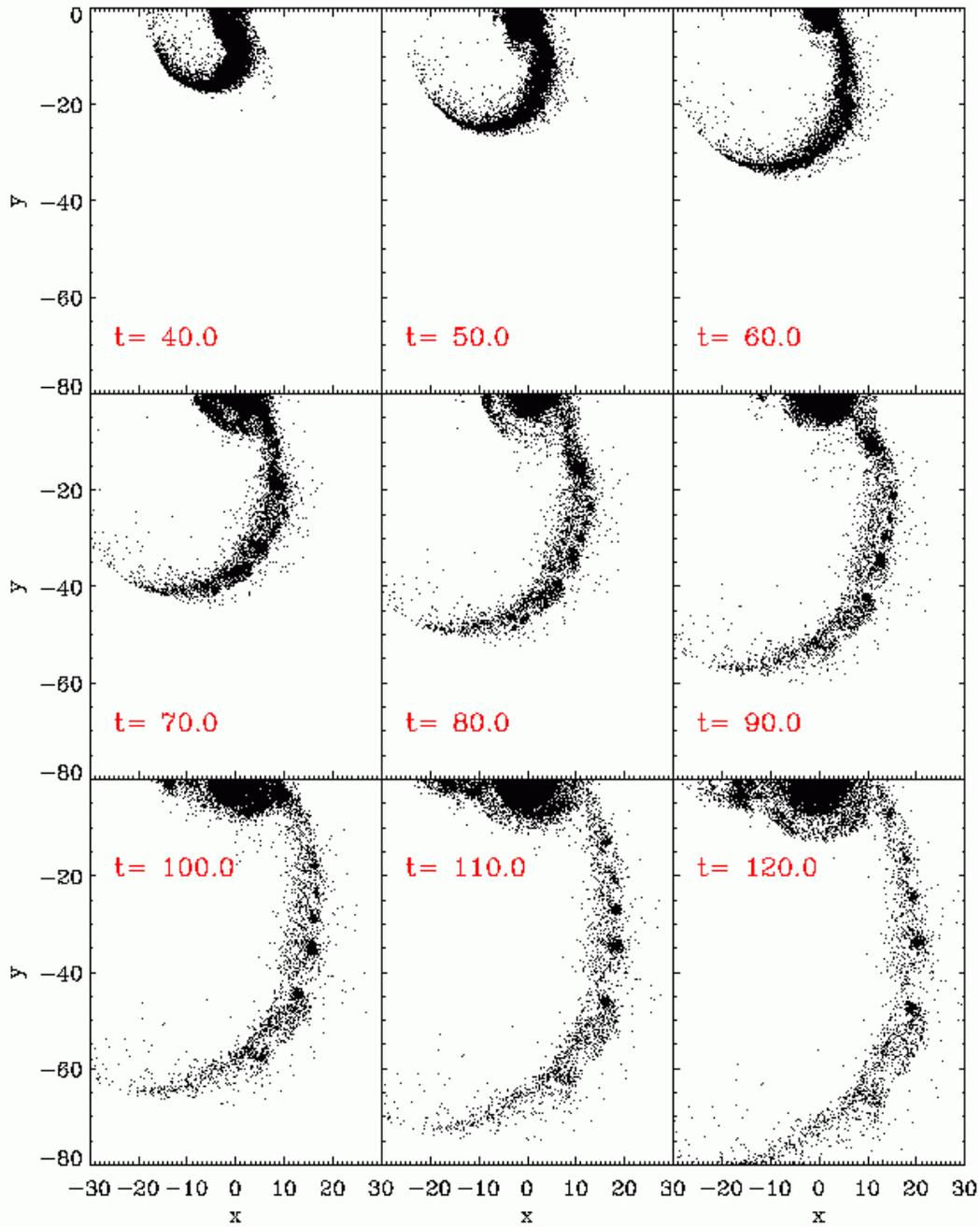,width=14cm}
        \caption{Time evolution of stellar component in model EG2, projected on
               the orbital plane. Several collapsed objects form.}
   \label{fig_11G320KQHG3STH}
   \end{center}
\end{figure*}
\begin{figure*}
  \begin{center}
    \leavevmode
    \epsfig{file=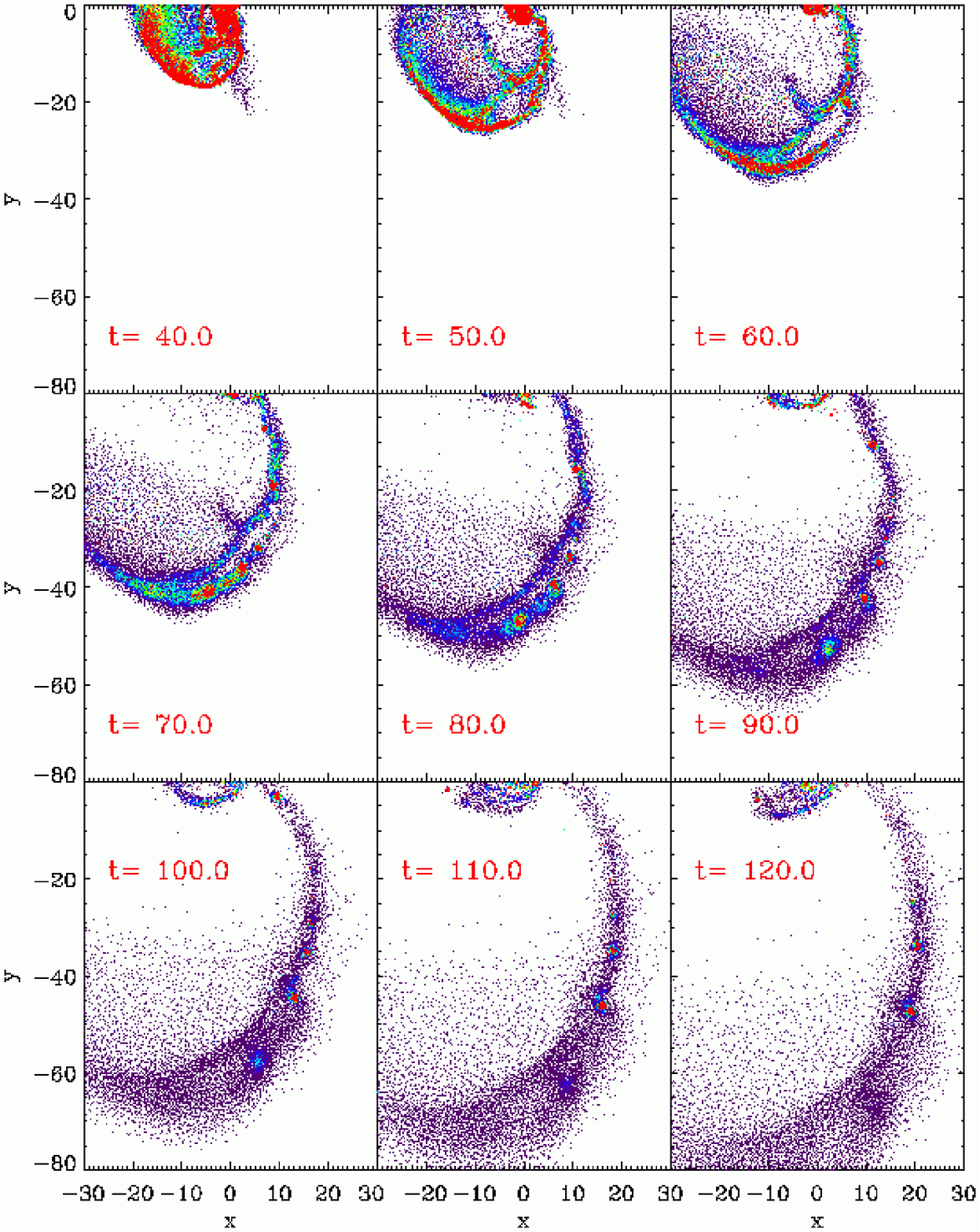,width=14cm}
        \caption{Time evolution of gas component in model EG2, projected on
               the orbital plane. The logarithmic gas densities are color
               coded. Several collapsed objects form.}   
   \label{fig_11G320KQHG3STH_gas}
   \end{center}
\end{figure*}

We have run a set of four models including gas, which is distributed in an
exponential disc. Our models comprise a 
combination of discs which have the same scale length as the stellar
disc or a scale length which is a factor of $3$ larger. The mass of
the gaseous discs is either $10\%$ or $30\%$ of the total disc mass, which
is fixed at $m_d=1$ as before in model C. The scale height of the gaseous disk
is $z_{0,g}=0.02$. The resulting combinations
are listed in Table \ref{tab_gas}. The less massive discs in models G1
and EG1 have been resolved with 32000 SPH particles, while the more
massive ones in models G2 and EG2 are represented with 45000 SPH
particles. For the setup with the massive, extended gas disc in model EG2, we
have matched the surface density of the gas at the 
solar radius to the observations of \citet{dame1993}, adopting a value of
$\Sigma_{r_\odot}=10.2$ M$_\odot$ pc$^{-2}$. Scaling our galaxy to the Milky
Way, we fix the corresponding surface density in our gas disc
accordingly. Model EG2 hence is more likely to resemble the
actual gas distribution in late type galaxies. Nevertheless, it is important
to study the possible effects of different gas distributions in the progenitor
disc on the distribution of gas in the tidal arm. The initial surface
densities of the gas components in our models are shown in Figure
\ref{fig_gas_surf}. 

In all models, the gas has been evolved with an isothermal
equation of state at temperature $T=12000 K$. The functional
form of the gravitational softening of the SPH particles is the same
spline kernel used for the $N$-body particles. The gravitational
softening length of the SPH particles is always identical to their SPH
smoothing length. This approach prevents effects of artificial
numerical collapse or stabilization against such
\citep{bate_burkert97}. In addition, we use the gravitational softening length
of the stellar disc particles as a lower limit on the softening length of the
SPH particles. Note that this can only lead to underestimating the degree of
collapse and not to artificially enhancing it.
The remaining parameters of the simulation
are as described in Section \ref{sec_params}. 

In order to demonstrate the stability of the models, we evolve the high
resolution galaxy in model EG2 in isolation. In Figure \ref{fig_qprof} we show
the initial radial profile of the Toomre stability parameter Q
\citep{toomre1964} for the gas, modified for a composite disc made out of gas
and stars \citep{wang1994,naab_ostrik2006}:
\beq
 Q_{g}=\f{\kappa \sigma_g}{\pi G \Sigma_g} \left(1+\f{\Sigma_*
     \sigma_g}{\Sigma_g \sigma_*} \right) \,.
\eeq
Here, $\kappa$ is the epicyclic frequency, $\sigma$ is velocity dispersion
and $\Sigma$ the surface density, indices $g$ and $*$ refer to gas and stars,
respectively.
Between 6 and 10 kpc, the disc is only marginally stable, but as we impose a
lower limit on the gravitational softening of the gas particles, this slightly
stabilizes the disc. We observe no local collapse during the evolution of the
disc. The rotation 
curves and surface density profiles are presented in Figure
\ref{fig_stability}. We show the profiles for $t=25=2.14 T_{1/2}$ and
$t=45=3.85 T_{1/2}$ where $T_{1/2}$ is 
the rotation period at the half-mass radius. When the galaxy is scaled to the
Milky Way, this corresponds to $t=327.5$ Myr and $t=589.5$ Myr,
respectively. The galaxy remains globally stable for these time
scales. Note that the first encounter of the two galaxies
takes place at around $t\approx25$, so by this time the perturbation due to
the tidal interaction will be much stronger than any effects internal to the
galaxy.

In Figure \ref{fig_rho_gas} we show the stellar densities along the tidal tails
in all four gas models. Model C is shown as well to allow for an easy
comparison. The densities plotted in Figure \ref{fig_rho_gas} are
obtained in the same way as those plotted in Figures
\ref{fig_rho_normsoft}-\ref{fig_rho_disc_halo}. Note that only stellar
material and no gas was taken into account. Only models G2 and EG2
have developed collapsed structures in the tidal tail at $t=120$. The
central stellar densities inside those collapsed objects is higher
than in the low resolution pure $N$-body models A and B. Those failed
to reach densities similar to observed dwarf galaxies (shaded region in
Figure \ref{fig_rho_gas}), while the objects in G2 and EG2 easily reach
such central densities. To illustrate the formation sequence of these
objects, we show the evolution of the tidal tail in model G1 in
Figures \ref{fig_11G320KQHG1STH}-\ref{fig_11G320KQHG1STH_gas} and that
of model EG2 in Figures
\ref{fig_11G320KQHG3STH}-\ref{fig_11G320KQHG3STH_gas}. The stellar
material is shown in the first figures, i.e. \ref{fig_11G320KQHG1STH}
for G1 and \ref{fig_11G320KQHG3STH} for EG2, while the evolution of the
gaseous component is shown in Figure \ref{fig_11G320KQHG1STH_gas} for
G1 and \ref{fig_11G320KQHG3STH_gas} for EG2. 

In model G1, the gas is ejected from the progenitor
disc together with the surrounding stellar material. It is then
compressed into a thin line along the middle of the tidal tail. A
significant fraction of the gas in the tidal tail flows back onto the
remnant galaxy. Those regions in the tail which have initially high
densities (see panel for t=50 in Figure \ref{fig_11G320KQHG1STH_gas})
are dispersed by the expanding tail so that at late stages (t=120) no
density enhancements have survived in the tidal tail. However, in
model EG2, the more extended setup of the gas disc leads to much more
gas which is tidally ejected from the progenitor disc. Later in the
evolution of the tidal tail in model EG2, the gas forms a transient
bifurcated structure (panels at $t=50,... 90$ in Figure
\ref{fig_11G320KQHG3STH_gas}). The gaseous tidal tail extends about
$\Delta r=20$ further than the stellar tidal tail. Both these features
in the gas distribution are also present in observed tidal tails \citep[see
  e.g.][]{hibbard1994,hibbard1999,hibbard2000,hibbard2001}. At later
stages the tidal tail in model EG2 forms collapsing objects ($t\ge 80$)
in the mid region of the tail, in contrast to the pure $N$-body model
C and model G1.

Model EG2 forms at least 5 collapsed objects in the tidal tail, while in model
G1 there exists none. Looking at the distribution of the objects in EG2 more
closely, they are slightly offset from the center line of the tidal tail
towards its leading outer edge. This is a result of the offset between stellar
and gaseous material in the tail, which develops as a consequence of the
different initial mass distributions and the dissipative nature of the gas
\citep[see][ for details]{mihos2001}. Apparently, the gas triggers local
gravitational collapse in the tidal tail where its own density is maximal, as
one would expect for gravitational instabilities in a gas distribution. For
the overall tidal tail, this leads to the collapse of both the stellar and
gaseous components slightly offset from the center line of the stellar
tail. After the gas has triggered the collapse, it settles in the centers of
the objects formed. As our model does not include a numerical recipe for star
formation and associated feedback, the gas forms very dense knots in the
center of each stellar body. 

Note that not only models G2 and EG2 formed collapsed objects. Also in
model EG1, there forms a massive collapsed object. However, the object
is located in the inner part of the tidal tail and the object falls
back into the remnant galaxy where it is tidally disrupted. In the
late phase of the tidal tail, as shown in Figure \ref{fig_rho_gas},
the object has long disappeared. So it would be wrong to believe that
massive gas discs in the progenitor galaxy, like in models G2 and EG2,
are required to form TDGs. Also a less massive gas disc can lead to
their formation, provided that the disc is extended enough. Also, the
mere presence of a dissipational component does not lead to the
formation of collapsed objects, as clearly indicated by model G1.

\begin{figure}
  \begin{center}
    \leavevmode
    \epsfig{file=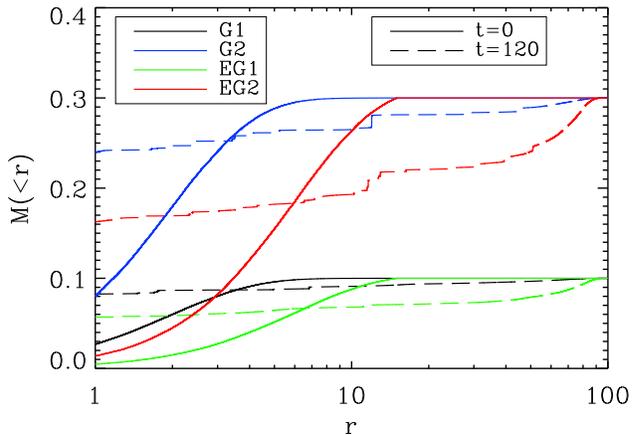,width=8.5cm}
        \caption{Cumulative gas mass for models G1, G2, EG1 and EG2 at t=0
    (solid lines) and at t=120 (dashed lines)}     
   \label{fig_cumgas}
   \end{center}
\end{figure}

In order to quantify how much gas is torn out of the progenitor disc, we plot
radial profiles of the cumulative gas mass for all models, see 
Figure \ref{fig_cumgas}. The profiles of the initial conditions are
shown as well as the profiles at $t=120$, the same evolved state of the tidal
arm which is also shown in Figure \ref{fig_rho_gas}. In model G1, a mass of 
$m_g \approx 0.09$ in gas is contained within the remnant and its immediate
surroundings ($r<5$). This corresponds to $90\%$ of the total gas
mass in this model. The distribution of the gas in the tidal tail is very
smooth, no gaseous substructures are present in G1. A similar smooth
distribution can be seen in model EG1, although more mass is deposited further
out in the tidal arm due to the more extended initial distribution. In model
G2, the mass of the gas inside and close to the remnant is $\approx 0.25$ while
the remaining gas with a mass of $\approx 0.05$ is located outside. Assuming an
inner cut for the tidal tail of $r=20$, the gas mass inside the tidal tail is
$m_{g,tail} \approx 0.03$, more than a factor of $3$ more than in model
G1. This clearly 
shows the effect of the gas distribution in the progenitor disc, as both G1
and EG1 have the same total mass and only differ in the corresponding
distribution. In model EG2, much more gas is torn out of the progenitor disc
than in any other of our gas models, as one would expect from the massive
\ti{and} extended disc in the progenitor. A mass of around $0.12$ is located
at 
$r>5$, which corresponds to $40\%$ of the total gas mass. Assuming again an
inner cut for the tidal tail of $r=20$, we find that a mass of
$\approx 0.08$ resides outside $r=20$, corresponding to $27\%$ of the total
gas mass. Some of the density peaks in Figure \ref{fig_rho_gas} can also
be found as jumps of the cumulative mass at the location of the object. 

We conclude that in contradiction to the previous findings of
\citet{barnes_h1996}, gas is the most important factor in the formation of
collapsed objects inside the tidal tails of merging galaxies. Also the
processes which lead to the formation of the offset between stellar and
gaseous tails are in the end not responsible for the suppression of TDG
candidate objects, as suggested by \citet{hibbard1999}, but in the contrary
are closely linked to their formation. The gas disc in the progenitor galaxy
needs to supply enough gas into the tidal arm to trigger the onset of
collapse there. The mere presence of a dissipative component is not enough. 

\subsection{Properties of a TDG Candidate}
\label{sec_tdg}
\begin{figure}
  \begin{center}
    \leavevmode
    \epsfig{file=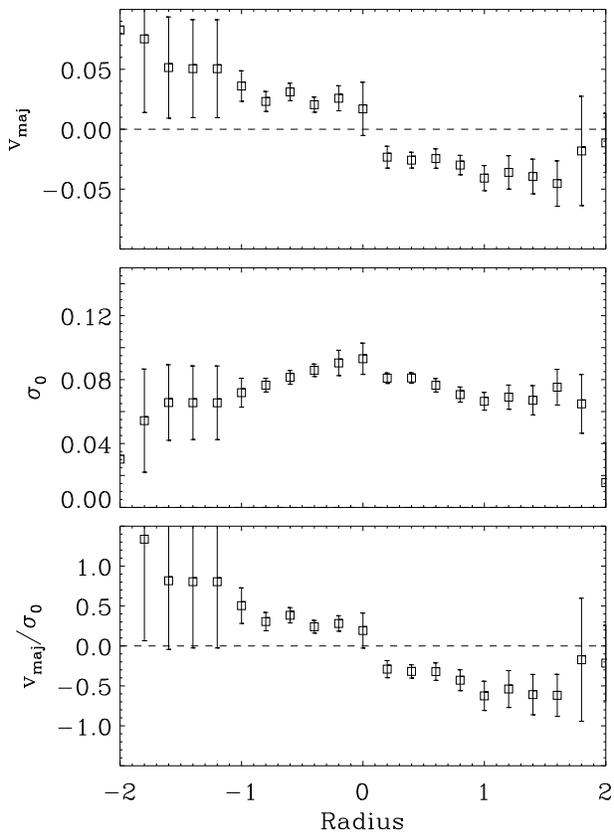,width=8.5cm}
        \caption{Observable major axis rotation velocity (upper panel) and
        velocity dispersion (mid panel) profiles of the most massive
        TDG candidate, plotting gas and stars. The lower panel shows the ratio
        of rotation to dispersion. For details see text.}
   \label{fig_dwarf}
   \end{center}
\end{figure}

\begin{figure}
  \begin{center}
    \leavevmode
    \epsfig{file=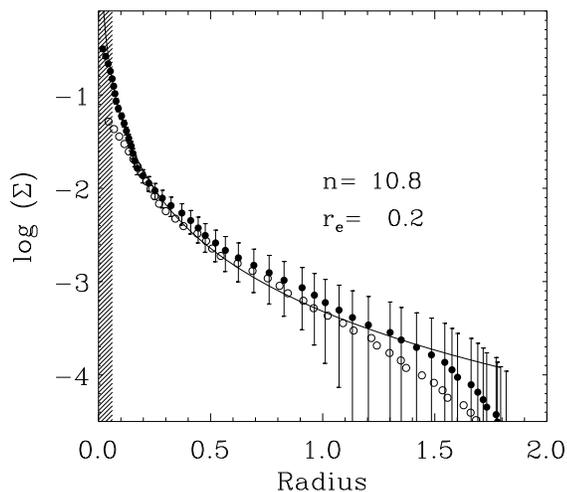,width=8.5cm}
        \caption{Surface brightness profile of the most massive TDG
        candidate. The shaded region corresponds to twice the gravitational
        softening radius. Filled circles represent gas plus stars, while empty
        circles are for the old stellar population only.}
   \label{fig_surf_dwarf}
   \end{center}
\end{figure}

The most massive TDG candidate object formed in our model EG2, located at
(20,-48) in the $t=120$ panel of Figure \ref{fig_11G320KQHG3STH_gas}, has a
mass of $0.062$, corresponding to $\approx 3.5 \times 10^8$
M$_\odot$ if the progenitor galaxy is scaled to the Milky Way. About
$70\%$ of its mass is in gas while there is no dark 
matter inside the object. The velocity dispersion of the
dark matter halo is too high to allow for their efficient
capture in the substructures of the tidal arm.

The object described above is our best candidate
object for a TDG. However, if this system resembles observed
substructures in tidal tails requires a more detailed study which
infers directly observable quantities. In order to assess the
properties of our candidate object, we convert the particle
distribution into an artificial image which is then analyzed in
a similar way as observations to avoid systematic differences. For the
details of the procedure, see \citet{naab2003}. 

We find an ellipticity of our
object of $\epsilon \approx 0.4$. In Figure \ref{fig_dwarf}, we show profiles
of the rotation velocity, the velocity dispersion and the ratio of the two
along the major axis. We plot gas and stars together, which is equivalent to
assuming that all gas was converted into stars after the TDG has
formed. Although there is a signature of rotation as 
well as a peak in the velocity dispersion, the magnitude of these are
rather low: A maximum rotation velocity of $0.05$ corresponds to
$13.1$ km s$^{-1}$ for a progenitor like the Milky Way, while the
maximum dispersion of $0.095$ corresponds to $24.9$ km s$^{-1}$. Such
values might be missed in observations, especially when
the object is possibly seen in less favorable projections.

We obtained a stellar surface density profile of our TDG candidate, see Figure
\ref{fig_surf_dwarf}, following the method used by
\citet{naab&trujilo2006}. The line represents the best fitting Sersic model,
corresponding to an index of $n=10.8$ and an effective radius of
$r_e=0.2$, which corresponds to $0.7$ kpc if the progenitor galaxy is scaled
to the Milky Way. The high index is caused by the concentration of the gas
towards the center. For the profile without the newly formed stars (open
circles in Figure \ref{fig_surf_dwarf}), we obtain index of $n\approx2$ and an
effective radius of $r_e=0.4$, corresponding to $1.4$ kpc. 

The stellar
properties of our TDG candidate are similar to observed properties of local
group dwarf galaxies \citep{mateo1998} as well as dwarf galaxies in groups or
clusters \citep{jerjen2000, baraza2003, derijcke2005}. But given the
uncertainty in the evolution of the gas in the shallow potential well, a
more realistic model for star formation might well affect the results. We
expect that values for $n$ and $r_e$ with and without gas outline the most
extreme cases.  


\section{SUMMARY AND DISCUSSION}
\label{sec_conclusion}

In this study we investigated the formation of tidal dwarf galaxies. Since the
early simulations of \citet{barnes_hern1992} and \citet{elmegr1993}, not much
work has been done regarding the detailed study of TDG formation. A well
accepted model for the formation of TDGs has, up to now, still to be
established. We have studied the importance of resolution and force softening
on the results of numerical simulations of tidal tails. Although particle
noise has already been suggested by \citet{barnes_hern1992} as an important
effect in this respect, no study prior to this work had investigated in detail
the effects of different resolutions on TDG formation. Performing an
extensive resolution study of pure $N$-body simulations, ranging from a total
of $288000$ particles to $4.128 \times 10^6$ particles, we find that 
with increasing resolution the collapse of objects inside the tidal
tail is subsequently suppressed. At a resolution level of $640000$
particles in the progenitor stellar disc, the tidal tail only develops
overdense regions which do not collapse within $200$ code time units,
corresponding to $\Delta t=2.62$ Gyr if the progenitor galaxy is
scaled to the Milky Way. This result is robust at even higher
resolution. The collapsed objects formed in the low resolution models cannot
be identified with dwarf galaxies, either, as they don't reach central
densities comparable to those of dwarf galaxies. For the initial conditions
presented here, we have shown that low resolution disk mergers lead to
collapsed objects in the tidal arms whereas for higher resolution this
collapse is suppressed.

In order to assess the effects of force softening on such simulations, we have
run a second set of simulations in which the gravitational softening length
has been varied. Using these models, we have shown that the lack of dense,
collapsed objects in the tidal tails  for high enough particle numbers is not
affected by our particular choice of softening lengths. 
Furthermore we have demonstrated that the resolution of the stellar
disc of the progenitor galaxy seems to be most important to suppress the
artificial collapse of regions in the tidal tail. The resolution of the
corresponding dark matter halo only plays a minor role in preventing
artificially collapsing objects. We hence conclude that whether or not we form
collapsed objects mainly depends on the resolution of the initial disk.
The previous findings of \citet{barnes_h1996} might originate from their low
nuemrical resolution.
Based on this study, we suggest the
use of at least half a million particles for modeling the progenitor disc and
twice that number for the dark matter halo in future simulations of tidal dwarf
galaxy formation. 

We have studied the effects of gas in tidal tails using
four different gas distributions in the progenitor discs. The presence of a
dissipative component alone is not sufficient for the formation of TDG like
objects. However, if the gas reservoir in the progenitor disc is massive
and / or extended enough, the tidal tail contains enough gas to trigger the
collapse of objects. The TDG candidates in our gaseous models easily
reach central densities comparable to dwarf galaxies, in contrast to
their counterparts in low resolution pure $N$-body simulations. Our findings
favor a dissipation supported scenario for TDG formation, as suggested by
\citet{elmegr1993} and \citet{kaufman1994}.

We have analyzed the most massive TDG candidate object in greater
detail. It has a mass of $\approx 3.5 \times 10^8$
M$_\odot$ if the progenitor galaxy is scaled to the Milky Way. The
object contains no dark matter and $70\%$ of its mass is in
gas. There exists a clear signature of rotation but the maximum
rotation velocity is $13.1$ km s$^{-1}$. The velocity dispersion profile shows
a clear peak in the 
center, corresponding to $24.9$ km s$^{-1}$. The surface density of
our candidate object can be reasonably fit with a Sersic profile of
index $n=10.8$ and effective radius of $r_e=0.7$ kpc. If we analyze only the
stellar component, we find $n\approx2$ and 
$r_e=1.4$ kpc. Our values for $n$ and $r_e$ outline
extreme cases, in which either no or all gas is transformed into stars. A
detailed model for star formation might result in values somewhere in
between. Stellar feedback could also expel some fraction of the gas from the
system. We will address these interesting questions in future simulations.
Given the uncertainties in the treatment of gas physics and star
formation, we find reasonable agreement with observations.

Our results demonstrate the importance of the gas component and the need for
high resolution simulations in order to study the formation of TDGs in
mergers. In this context, models should ideally also take star formation and
stellar feedback into account. Such simulations including
additional physical processes are beyond the scope of this paper,
which aimed at studying more fundamental effects, including numerical
resolution and force softening. More detailed models which take star
formation and feedback into account will be investigated in future work.

\section*{Acknowledgments}

M.W. acknowledges support by the Volkswagen Foundation under grant I/80
040. The simulations have partly been performed at the Rechenzentrum of the Max
Planck Society, Garching, Germany. We thank the anonymous referee for his
valuable comments.

\label{lastpage}

\bibliographystyle{mn2e}
\bibliography{lit}

\end{document}